\documentclass[10pt, conference, letterpaper]{IEEEtran}
\usepackage[letterpaper, left = 0.68in, right =0.68in, top = 0.7in, bottom = 1in]{geometry}
\usepackage[noadjust]{cite}
\usepackage{amsmath,amssymb,amsfonts}
\usepackage{algorithmic}
\usepackage{graphicx}
\usepackage{textcomp}
\usepackage{mathtools}
\usepackage{adjustbox}
\usepackage{tikz}
\usetikzlibrary{arrows, fit, matrix, positioning, shapes, backgrounds}
\usepackage{xcolor}
\def\BibTeX{{\rm B\kern-.05em{\sc i\kern-.025em b}\kern-.08em
    T\kern-.1667em\lower.7ex\hbox{E}\kern-.125emX}}

\usepackage{supertabular}    
\usepackage{array} 
\usepackage{soul}
\usepackage{numprint}     
\usepackage{tikz}
\usepackage{makecell}   
\usepackage{enumitem}                              
\usepackage{longtable} 
\usepackage{wrapfig}
\usepackage{subcaption}
\usepackage{pgfplots}
\usepgfplotslibrary{groupplots}
\usepackage{booktabs}

\usepackage{amsthm}
\usepackage{amsmath, amssymb}

\pgfplotsset{vasymptote/.style={
    before end axis/.append code={
        \draw[densely dashed, line width=1.5pt] ({rel axis cs:0,0} -| {axis cs:#1,0})
        -- ({rel axis cs:0,1} -| {axis cs:#1,0});
    }
}}

\DeclarePairedDelimiter\ceil{\lceil}{\rceil}
\DeclarePairedDelimiter\floor{\lfloor}{\rfloor}

\newtheorem{theorem}{Theorem}

\usepackage[utf8]{inputenc}
\usepackage[acronym]{glossaries}
\usepackage[ruled]{algorithm2e}

\newcommand{\fakepar}[1]{\vspace{1mm}\noindent \textbullet \hspace{1mm}\textit{#1.}}   

\newcommand{\egc}{e.\,g., }
\newcommand{\iec}{i.\,e., }

\newcommand{\wrt}{w.\,r.\,t.\ }

\begin{document}

\title{A Computationally Efficient 2D MUSIC Approach for 5G and 6G Sensing Networks}

\author{
    \IEEEauthorblockN{
        Marcus Henninger\IEEEauthorrefmark{1}\IEEEauthorrefmark{2},
        Silvio Mandelli\IEEEauthorrefmark{1},
        Maximilian Arnold\IEEEauthorrefmark{1},
        and Stephan ten Brink\IEEEauthorrefmark{2}
        }

	\IEEEauthorblockA{
	\IEEEauthorrefmark{1}Nokia Bell Labs, 70435 Stuttgart, Germany, \\
	\IEEEauthorrefmark{2}Institute of Telecommunications, University of Stuttgart, 70659 Stuttgart, Germany\\
	   E-mail: marcus.henninger@nokia.com, \{silvio.mandelli, maximilian.arnold\}@nokia-bell-labs.com}}

\maketitle

\newacronym{1D}{1D}{one-dimensional}
\newacronym{2D}{2D}{two-dimensional}
\newacronym{csi}{CSI}{channel state information}
\newacronym{ula}{ULA}{uniform linear array}
\newacronym{ofdm}{OFDM}{orthogonal frequency-division multiplexing}
\newacronym{awgn}{AWGN}{additive white Gaussian noise}
\newacronym{ev}{EV}{eigenvalue}
\newacronym{dd}{DD}{delay-Doppler}
\newacronym{tf}{TF}{time-frequency}
\newacronym{qpsk}{QPSK}{Quadrature phase-shift keying}
\newacronym{evd}{EVD}{eigenvector decomposition}
\newacronym{cfar}{CFAR}{constant false alarm rate}
\newacronym{music}{MUSIC}{MUltiple SIgnal Classification}
\newacronym{rmse}{RMSE}{root-mean-square error}
\newacronym{flop}{FLOP}{floating-point operation}
\newacronym{jcas}{JCAS}{Joint Communication and Sensing}
\newacronym{b5g}{B5G}{Beyond 5G}
\newacronym{awgn}{AWGN}{additive white Gaussian noise}
\newacronym[plural=AoAs,firstplural = angles of arrivals (AoAs)]{aoa}{AoA}{angle of arrival}
\newacronym{nr}{NR}{New Radio}
\newacronym{snr}{SNR}{signal-to-noise ratio}
\begin{abstract}
Future cellular networks are intended to have the ability to sense the environment by utilizing reflections of transmitted signals. Multi-dimensional sensing brings along the crucial advantage of being able to resort to multiple domains to resolve targets, enhancing detection capabilities compared to \gls{1D} estimation. However, estimating parameters jointly in 5G \glsentrylong{nr} systems poses the challenge of limiting the computational complexity while preserving a high resolution. To that end, we make use of \gls{csi} decimation for \gls{music}-based joint range-\glsentrylong{aoa} estimation. We further introduce multi-peak search routines to achieve additional detection capability improvements. Simulation results with \gls{ofdm} signals show that we attain higher detection probabilities for closely spaced targets than with \gls{1D} range-only estimation. Moreover, we demonstrate that for our considered 5G setup, we are able to significantly reduce the required number of computations due to \gls{csi} decimation.
\end{abstract}

\vspace{0.25cm}


\glsresetall
\makeatletter{\renewcommand*{\@makefnmark}{}
\footnotetext{\copyright\;2022 IEEE. Personal use of this material is permitted. Permission from IEEE must be obtained for all other uses, in any current or future media, including reprinting/republishing this material for advertising or promotional purposes, creating new collective works, for resale or redistribution to servers or lists, or reuse of any copyrighted component of this work in other works.}\makeatother}

\section{Introduction}\label{sec:intro}

\gls{jcas} is expected to be one of the main features of \gls{b5g} and 6G cellular networks and is currently attracting a lot of attention in the research community \cite{jcas_vde}. Future networks should be able to extract information about the physical world from the channel, performing de-facto sensing operations. Road user protection \cite{5GCAR} and positioning in indoor scenarios \cite{ToA_AoA_Loc} are two of the various sensing applications. In communication systems, the \gls{csi} is estimated as part of the necessary steps for communication purposes. \Gls{ofdm} radar \cite{ofdm_concept} can be used to exploit this information by means of the reflections from objects illuminated by the transmitted signal to estimate their ranges and \glspl{aoa}. Among possible estimation techniques, we consider the \gls{music} algorithm  \cite{MUSIC_original}, as it offers super-resolution capabilites over the Periodogram algorithm and allows coping with generic antenna array shapes.

Estimating range and \gls{aoa} jointly offers the major advantage that targets can be resolved in two dimensions and consequently improves the ability to detect closely spaced targets compared to \gls{1D} estimation. In general, joint estimation of multiple parameters is a well-researched field and has been discussed \egc in \cite{JADE} and \cite{Joint_UWB}. For the particular case of jointly estimating range and \gls{aoa} with \gls{music} and \gls{ofdm} signals, \textit{SpotFi} \cite{spotfi} has made use of the spatial smoothing technique~\cite{smoothing_orig}. Differently from \cite{spotfi}, which deals with WiFi signals, we consider 5G \gls{nr} numerologies \cite{5G_3gpp}, requiring a large number of subcarriers to achieve the necessary bandwidth for a high range resolution. The resulting increased \gls{csi} matrix dimension would lead to a complexity that is too high to be handled in real-time applications.
\\\indent To that end, we enhance the spatial smoothing technique by properly decimating the estimated \gls{csi}, adopting the approach used in \cite{MUSIC_range_doppler} for joint range-Doppler estimation in the context of automotive radars. Our proposal enables high resolution in 5G sensing applications while drastically reducing the computational complexity. We also formulate a necessary condition for detecting multiple targets with multi-dimensional spatial smoothing. Similar to the motivation of earlier works (\unskip \cite{gold_MUSIC, 2D_MUSIC_realtime}), we want to avoid an inefficient grid search in the resulting \gls{2D} \gls{music} spectrum. We therefore present a peak search routine using Powell's algorithm \cite{powell} that can detect multiple targets at once.
Moreover, we can reduce the missed detection probability by coherently removing the contribution of previously detected targets utilizing subspace tracking methods \cite{subspace_tracking}. All our algorithms can straightforwardly be extended to higher dimensions for estimating additional parameters and can directly be implemented on top of 5G \gls{nr} architectures. 

The paper is structured as follows: after defining the signal model in Section~\ref{sec:sig_model}, we present our algorithms in Section~\ref{sec:algo}. Section~\ref{sec:results} introduces simulation scenario and assumptions and discusses our method's benefits based on the results. The paper is wrapped up with a conclusion (Section~\ref{sec:conclusion}).

\section{Signal Model}\label{sec:sig_model}

We consider a single TX antenna illuminating the environment by transmitting phase-modulated constant envelope complex symbols $\mathbf{s} =\begin{bmatrix}{s}_{1},\ {s}_{2},\ \hdots,\ {s}_{N} \end{bmatrix}$, modulated onto $\mathit{N}$ subcarriers of an \gls{ofdm} signal with subcarrier spacing $\Delta f$. The carrier wavelength is denoted by $\lambda$. 
While we only investigate static targets in this paper, the extension to Doppler estimation could be accomplished by considering multiple \gls{ofdm} symbols. 
\\The RX \gls{ula} is co-located with the transmitter and comprises $\mathit{K}$ antennas with element spacing $\mathit{d}$. Let the received symbols at array element \textit{k} be written as a row vector $\mathbf{y}_{k}=\begin{bmatrix}
{y}_{k,1},\ {y}_{k,2},\ \hdots,\ {y}_{k,N} \end{bmatrix}$.
Assuming knowledge of the TX symbols, the frequency domain \gls{csi} at each RX antenna is obtained by carrying out an element-wise division
\begin{equation}
c_{k, n} = \frac{y_{k, n}}{s_{n}}.
\label{eq:radar}	
\end{equation} 
Stacking the \gls{csi} of all antenna elements vertically yields the $\mathit{K \times N}$ \gls{csi} matrix $\mathbf{C} = \begin{bmatrix} 
\textbf{c}_{1},\ \textbf{c}_{2},\ \hdots,\ \textbf{c}_{K} \end{bmatrix}^\text{T}$. Assuming $Q$ impulsive scatterers as targets, each of them generates a reflection according to azimuth \gls{aoa} $\theta_q$ and range $r_q = \frac{\tau_{q}c}{2}$, where $\tau_q$ is the delay of the reflection caused by the $\mathit{q}$-th target and $\mathit{c}$ the speed of light. The \gls{csi} matrix can be expressed as the superposition of the channel contributions of all $\mathit{Q}$ targets
\begin{equation}
\mathbf{C} = \sum_{q=1}^{Q} h_{q}\mathbf{b}(\theta_q)\mathbf{a}(r_q)^\text{T} + \textbf{Z},
\label{eq:csi}	
\end{equation} 
where $\mathit{h_{q}}$ is the complex coefficient of the $\mathit{q}$-th target and $\mathbf{Z}$ the $\mathit{K \times N}$ random complex \gls{awgn} matrix. Note that we neglect the influence of self-interference, as it is not the scope of this work and has previously been addressed in literature \cite{selfinterference}. Assuming that $N\Delta f \ll f_c$ (carrier frequency) and $r_q \gg Kd$, the vectors $\mathbf{b}(\theta_q)$ and $\mathbf{a}(\mathit{r_{q}})$ are given as
\begin{align}
\mathbf{b}(\theta_q) &= \begin{bmatrix}
		1, \ e^{j2\pi  \frac{d}{\lambda} \sin(\theta_{q})},\ \dots, \ e^{j2\pi (K -1) \frac{d}{\lambda} \sin(\theta_{q}) }
\end{bmatrix}^\text{T}  \\
\mathbf{a}(r_{q}) &= \begin{bmatrix}
       1, \ e^{-j2\pi \Delta f \cdot 2\frac{r_{q}}{c}}, \ \dots, \ e^{-j2\pi (N - 1) \Delta f \cdot 2\frac{r_{q}}{c}}
\end{bmatrix}^\text{T} 
\label{eq:channel_vectors}	
\end{align} 
and represent the linear phase shifts in the antenna and subcarrier dimension induced by $\mathit\theta_q$ and $\mathit{r}_{q}$ of the $\mathit{q}$-th target.
\section{Proposed Algorithm}\label{sec:algo}

\subsection{Spatial Smoothing with \gls{csi} Decimation}\label{sec:csi_processing}

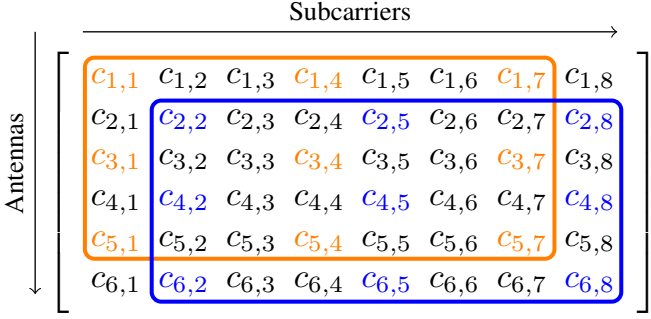
\begin{figure}[t]
	\resizebox{9cm}{!}{\begin{tikzpicture}

\tikzstyle{boxphaseI} = [draw,rounded corners=.1cm,inner sep=5pt,minimum height=6em, text width=13em, align=left,very thick]

\newcommand*\circled[1]{\tikz[baseline=(char.base)]{
            \node[shape=circle,draw,inner sep=2pt] (char) {#1};}}

\matrix (A) [
matrix of math nodes,
anchor = north west,
inner sep = 0.25em,
left delimiter={[},
right delimiter={]}
]
{ 
  \color{orange}{c_{1,1}} & c_{1,2} & c_{1,3} & \color{orange}{c_{1,4}} & c_{1,5} & c_{1,6} & \color{orange}{c_{1,7}} & c_{1,8}\\
  c_{2,1} & \color{blue}{c_{2,2}} & c_{2,3}  & c_{2,4} & \color{blue}{c_{2,5}} & c_{2,6} & c_{2,7} & \color{blue}{c_{2,8}}\\
  \color{orange}{c_{3,1}} & c_{3,2} & c_{3,3}  & \color{orange}{c_{3,4}} & c_{3,5} & c_{3,6} & \color{orange}{c_{3,7}} & c_{3,8}\\
  c_{4,1} & \color{blue}{c_{4,2}} & c_{4,3}  & c_{4,4} & \color{blue}{c_{4,5}} & c_{4,6} & c_{4,7} & \color{blue}{c_{4,8}}\\
  \color{orange}{c_{5,1}} & c_{5,2} & c_{5,3}  & \color{orange}{c_{5,4}} & c_{5,5} & c_{5,6} & \color{orange}{c_{5,7}} & c_{5,8}\\
  c_{6,1} & \color{blue}{c_{6,2}} & c_{6,3}  & c_{6,4} & \color{blue}{c_{6,5}} & c_{6,6} & c_{6,7} & \color{blue}{c_{6,8}}\\
};

\node[boxphaseI,orange, label={[orange, above]: \scriptsize}, anchor = north west] (b1) at (0.1,-0.05){};
\node[boxphaseI,blue, label={[orange, above]: \scriptsize}, anchor = north west] (b2) at (0.8,-0.5){};

\coordinate (arrow_1) at (0.1, 0.2) {};
\coordinate (arrow_2) at (5.7, 0.2) {};
\coordinate (arrow_3) at (-0.4, 0.2) {};
\coordinate (arrow_4) at (-0.4, -2.55) {};

\draw[->] (arrow_1) -- (arrow_2) node[midway, above] {\footnotesize Subcarriers};
\draw[->] (arrow_3) -- (arrow_4) node[midway, above, rotate = 90] {\footnotesize Antennas};

\end{tikzpicture}}
	 \caption{Illustration of two different sub-arrays obtained by proposed sub-array generation with apertures $\mathit{A_f} = 7$ and $\mathit{A_a} = 5$, decimations $\mathit{D_f} = 3$ and $\mathit{D_a} = 2$, and strides $\mathit{S_f} = \mathit{S_a} = 1$.}
	 \label{CSI_processing}
\end{figure}

Applying \gls{music} in two dimensions couples range and \gls{aoa} estimates to the same target and therefore improves the ability of discriminating different targets. However, previous work~\cite{smoothing_orig} states that for detecting multiple targets, a necessary condition is that the number of independent measurements for computing the sample covariance matrix $\hat{\mathbf{R}}_y$ must be greater or equal than $\mathit{Q}$. To achieve this with a single snapshot, \cite{spotfi} makes use of the spatial (in the algebraic sense) smoothing technique by generating $\mathit{L}$ sub-arrays $\begin{bmatrix}\mathbf{C}_{s, 1},\ \mathbf{C}_{s, 2},\ \hdots,\ \mathbf{C}_{s, L} \end{bmatrix}$ from  $\mathbf{C}$. The $\mathit{M \times L}$ smoothed \gls{csi} matrix $\tilde{\mathbf{C}}$ is then constructed as 
\begin{align}
\tilde{\mathbf{C}} &= \begin{bmatrix} 
    \text{vec}(\mathbf{C}_{s, 1}),\ \text{vec}(\mathbf{C}_{s, 2}),\ \hdots, \    \text{vec}(\mathbf{C}_{s, L})
    \end{bmatrix} \nonumber \\
&= \begin{bmatrix} 
    \mathbf{c}_{s, 1},\ \mathbf{c}_{s, 2},\ \hdots, \  \mathbf{c}_{s, L}
    \end{bmatrix},
\label{eq:CSI_smoothed}
\end{align}
with $\mathit{M}$ being the number of samples per sub-array and vec($\cdot$) the vectorization operator applied in row-major order. As an example, \textit{all} elements in the orange and blue boxes of Fig.~\ref{CSI_processing} form two independent sub-arrays. The sample covariance matrix can be computed as
\begin{align}
\hat{\mathbf{R}}_y = \frac{1}{M}\tilde{\mathbf{C}}\tilde{\mathbf{C}}^\text{H}.
\label{eq:corr_matrix}
\end{align}
After performing the \gls{evd} of $\hat{\mathbf{R}}_y$, the eigenvectors are partitioned into the $M \times Q$ signal subspace $\mathbf{U}_S$ corresponding to the $\mathit{Q}$ strongest eigenvalues and the complementary $M \times (M - Q)$ noise subspace $\mathbf{U}_N$. For estimating $Q$, the minimum description length method \cite{MDL} is used. The 2D \gls{music} spectrum is obtained by computing 
\begin{equation}
P_{\text{MU}}(r, \theta) = \frac{1}{(\mathbf{b}(\theta) \otimes \mathbf{a}(r))^\text{H} \textbf{U}_{N}\textbf{U}_{N}^\text{H}(\mathbf{b}(\theta) \otimes\mathbf{a}(r))},
\label{eq:MUSIC}
\end{equation}
where $\mathbf{b}(\theta)$ and $\mathbf{a}(\mathit{r})$ are the steering vectors for the trial range-azimuth pair $(\mathit{r}, \theta)$ and $\otimes$ is the Kronecker product.
\\Preserving the domain apertures of $\mathbf{C}$ leads to the dimensions of the sub-arrays growing rapidly, rendering the approach computationally expensive. Obviously, one could simply consider small sub-arrays and limit the complexity in this way. However, in modern 5G systems, even with $\Delta f$ = 60 kHz, the number of subcarriers $\mathit{N}$ could already exceed one thousand. To preserve the range resolution, it is prerequisite to have sub-arrays with a large frequency domain aperture $A_f$, as the range resolution $\Delta r$ of an \gls{ofdm} radar is
\begin{equation}
\Delta r = \frac{c}{2A_f \Delta f}.
\label{eq:range_res}	
\end{equation} 
Therefore, the range resolution is inversely proportional to the frequency aperture. With fixed $\Delta f$, one can only increase $A_f$, thus $M$, to improve ranging performance, quickly leading to computationally infeasible situations.

As introduced in \cite{MUSIC_range_doppler} for range-Doppler estimation, we decimate the sub-arrays, allowing us to keep a large aperture in both frequency (\iec resulting effective bandwidth) and spatial (\iec aperture of the antenna array) dimension. Besides apertures $A_f$ and $A_a$, we parametrize the generation of sub-arrays by defining decimation and stride between them, which shall be denoted as $\mathit{D}$ and $\mathit{S}$, respectively, with subscripts $\mathit{f}$ and $\mathit{a}$ representing frequency and spatial dimension. The antenna and subcarrier indices of the $\ell$-th sub-array of the smoothed \gls{csi} matrix $\tilde{\mathbf{C}}$ are denoted as vectors of length $M$
\begin{align}
\mathbf{a}_{\ell} &= \textbf{1}_{\tilde{N}_{f}} \otimes \begin{bmatrix}
		\tilde{a}_{\ell}, \ \tilde{a}_{\ell} + D_a, \ \dots, \ \tilde{a}_{\ell} + (\tilde{N}_{a} -1)D_a
\end{bmatrix}^{\text{T}}
\label{eq:ant_idxs} \\
\mathbf{f}_{\ell} &=  \begin{bmatrix}
		\tilde{f}_{\ell}, \ \tilde{f}_{\ell} + D_f, \ \dots, \ \tilde{f}_{\ell} + (\tilde{N}_{f}-1)D_f
\end{bmatrix}^{\text{T}} \otimes \textbf{1}_{\tilde{N}_{a}}
\label{eq:freq_idxs}
\end{align} 
to be used to sample $\mathbf{c}_{s, \ell}$ from $\mathbf{C}$ as 
\begin{align}
\mathbf{c}_{s,\ell} &= \begin{bmatrix} 
    \mathbf{C}_{a_{\ell, 1}, f_{\ell, 1}}, \ \mathbf{C}_{a_{\ell, 2}, f_{\ell, 2}}, \ \dots, \ \mathbf{C}_{a_{\ell, M}, f_{\ell, M}}
    \end{bmatrix}^\text{T},
\label{eq:c_sub_l}
\end{align}
where $\tilde{f}_{\ell}$ and $\tilde{a}_{\ell}$ are the respective initial indices, $\tilde{N}_{f} = \ceil*{A_{f}/D_{f}}$ and $\tilde{N}_{a} = \ceil*{A_{a}/D_{a}}$ the numbers of subcarriers and antennas per sub-array, and $\textbf{1}_{\tilde{N}_{f}}$ and $\textbf{1}_{\tilde{N}_{a}}$ all-ones vectors of length $\tilde{N}_{f}$ and  $\tilde{N}_{a}$. In total, $L = \tilde{S}_f\tilde{S}_{a}$ sub-arrays can be obtained, where $\tilde{S}_{f} = \floor*{(N - A_{f})/S_{f} + 1}$ and $\tilde{S}_{a} = \floor*{(K - A_{a})/S_{a} + 1}$ denote the number of different subcarrier and antenna sets as a result of striding. Referring to Fig. \ref{CSI_processing}, the initial full sub-arrays can be sampled by selecting only the \textit{colored} elements. After generating $\tilde{\mathbf{C}}$ in this way, the 2D \gls{music} spectrum is estimated using (\ref{eq:CSI_smoothed}) - (\ref{eq:MUSIC}), with the steering vectors being dependent on the sub-array definition parameters
\begin{align}
\tilde{\mathbf{b}}_{\phi_a}(\theta) &= \begin{bmatrix}
		1, \ e^{j\phi_a \sin(\theta)},\ \dots, \ e^{j \phi_a\tilde{N}_{a}\sin({\theta})}
\end{bmatrix}^\text{T}
\label{eq:steer_vec_angle} \\
\tilde{\mathbf{a}}_{\phi_f}(r) &= \begin{bmatrix}
       		1, \ e^{j \phi_f \cdot 2\frac{r}{c}},\ \dots, \ e^{j \phi_f \tilde{N}_{f} \cdot 2\frac{r}{c}}
\end{bmatrix}^\text{T} \label{eq:steer_vec_range},
\end{align}
where $\phi_a = 2\pi D_a \frac{d}{\lambda}$ and $\phi_f = -2\pi D_f \Delta f$ to ease notation. Due to decimation, the number of elements per sub-array $\mathit{M}$ is reduced from $A_f A_a$ to $\tilde{N}_f\tilde{N}_a$. Therefore, an equal frequency domain aperture, and thus resolution, can be achieved with sub-arrays with ca. $D_fD_a$ times less elements, drastically reducing the computational complexity. However, decimating in the subcarrier domain reduces the unambiguous range $r_{\text{max}} = \frac{c}{2 D_f \Delta f}$ by $\mathit{D_f}$. We account for this by choosing the ranges in (\ref{eq:steer_vec_range}) within $r_{\text{max}}$ to avoid aliasing.

Given that we want to discriminate $Q$ targets in a multi-dimensional space, at least $Q$ independent sub-arrays are necessary to compute $\hat{\mathbf{R}}_y$. If the targets are not resolvable in one domain, \egc they have the same range, we notice that striding only in the corresponding dimension, in that case subcarriers, does not generate independent measurements. We can then formalize a necessary condition for the separation of $Q$ targets in a multi-dimensional space.

\begin{theorem}\label{theo}
Consider signals generated by periodically sampling across $V$ different dimensions of interest, \egc subcarriers and antennas in Eq.~(\ref{eq:csi}). To separate $Q$ targets, that can be solved in at least one dimension of interest, with multi-dimensional spatial smoothing, at least $Q$ independent sub-arrays must be used with respect to \textbf{each dimension of interest}.
\end{theorem}

\begin{proof}
We provide a proof without considering additive noise, since the extension of the proof for noisy scenarios is straightforward. We need to separate $Q$ targets in a $V$-dimensional MUSIC spectrum. This means that the sample covariance matrix $\hat{\mathbf{R}}_y$ must be of rank $Q$, \iec it has $Q$ non-zero eigenvalues. We consider the worst-case scenario, that is that the $Q$ targets have exactly equal coordinates in $V-1$ dimensions. However, they can be discriminated in dimension $v$, \egc angle. Assume we have $Q' < Q$ independent sub-arrays, obtained by striding over dimension $v$, plus an arbitrary number of sub-arrays obtained by striding over the other $V-1$ dimensions. However, if the $Q$ targets have exactly the same coordinates over the $V-1$ dimensions, all the sub-arrays generated by striding over the $V-1$ dimensions will be equal to $\mathbf{q}$, that is one of the $Q'$ sub-arrays with only constant phase shifts, according to (\ref{eq:csi}). Therefore, when we use these sub-arrays to estimate the sample covariance matrix in (\ref{eq:corr_matrix}), their contribution will be the same of $\mathbf{q}$. This leads to a covariance matrix of rank $Q'$. Recalling that we need to have covariance matrices with at least $Q$ non-zero eigenvalues, we must have that $Q' \geq Q$.

Let us consider a 2D example with $Q$ targets at equal range $r$. According to (\ref{eq:csi})
\begin{equation*}
\mathbf{C}=\sum_{q=1}^Q h_q \mathbf{b}(\theta_q) \mathbf{a}(r)^\text{T} = \left( \sum_{q=1}^Q h_q \mathbf{b}(\theta_q) \right)\mathbf{a}(r)^\text{T} \;.
\end{equation*}
If we consider two sub-arrays $\mathbf{C}_{s,1}$ and $\mathbf{C}_{s,2}$ obtained by selecting the fist sub-array and by then striding of $S_f$ indices across subcarriers, respectively, we would have
\begin{align*}
\mathbf{C}_{s,1}&=\left(\sum_{q=1}^Q h_q \mathbf{\tilde{b}}(\theta_q) \right) \mathbf{\tilde{a}}(r)^\text{T} \;,
\\
\mathbf{C}_{s,2}&=\left(\sum_{q=1}^Q h_q \mathbf{\tilde{b}}(\theta_q) \right) \mathbf{\tilde{a}}(r)^\text{T} e^{-j2\pi S_f \Delta f \cdot 2 \frac{r}{c}}= \mathbf{C}_{s,1} e^{jW}\;,
\end{align*}
where the dependency on $\phi_a$ and $\phi_f$ has been dropped from the sub-array steering vectors defined in (\ref{eq:steer_vec_angle})-(\ref{eq:steer_vec_range}) to improve readability. The real phase constant is defined as $W=-2\pi S_f \Delta f \cdot 2 \frac{r}{c}$. Then, if these two sub-arrays are used to estimate $\hat{\mathbf{R}}_y$, one would have
\begin{align*}
\hat{\mathbf{R}}_y &= \frac{1}{2} \left( \mathbf{C}_{s,1} \mathbf{C}_{s,1}^\text{H} + \mathbf{C}_{s,2}\mathbf{C}_{s,2}^\text{H} \right) =
\\
&= \frac{1}{2} \mathbf{C}_{s,1} \mathbf{C}_{s,1}^\text{H} \left(1 + e^{jW}e^{-jW} \right) = \mathbf{C}_{s,1} \mathbf{C}_{s,1}^\text{H} \;.
\end{align*}
It is clear that the contribution of the two sub-arrays is completely dependent, thus they will contribute to generate a single eigenvalue. It follows that one would need to stride across the antenna dimension at least $Q$ times to generate $Q$ independent contributions, thus the necessary $Q$ eigenvalues to separate them in the angular domain. 
\end{proof}

\subsection{Peak Search Routine}

In order to find targets in the 2D space, the maxima of the \gls{music} spectrum defined in (\ref{eq:MUSIC}) need to be found. Avoiding an exhaustive grid search and allowing to detect multiple targets at once are the motives behind a proper peak search routine.
\\To find peaks in the spectrum, Powell's algorithm \cite{powell} implemented in Python's SciPy library is utilized. For the algorithm to converge to the 2D \gls{music} spectrum maxima, suitable starting points for the search are critical. For this, we first evaluate (\ref{eq:MUSIC}) in a coarse grid, where the sampling period in each dimension is equal to one half of the corresponding domain resolution. The locations of the $N_{\text{start}}$ highest \gls{music} spectrum values are selected as starting points. Each of them is used by Powell's algorithm to find a peak that is compared against a \gls{cfar} noise threshold $\gamma$ defined by a predefined probability of false alarm $\mathit{p_{FA}}$ \cite{braun_thesis}. If the peak value is below $\gamma$, it is discarded. 

\subsection{Coherent Target Cancelation}

\begin{figure}[h]
\centering
\begin{subfigure}[b]{0.5\textwidth}
	\includegraphics[width=\linewidth]{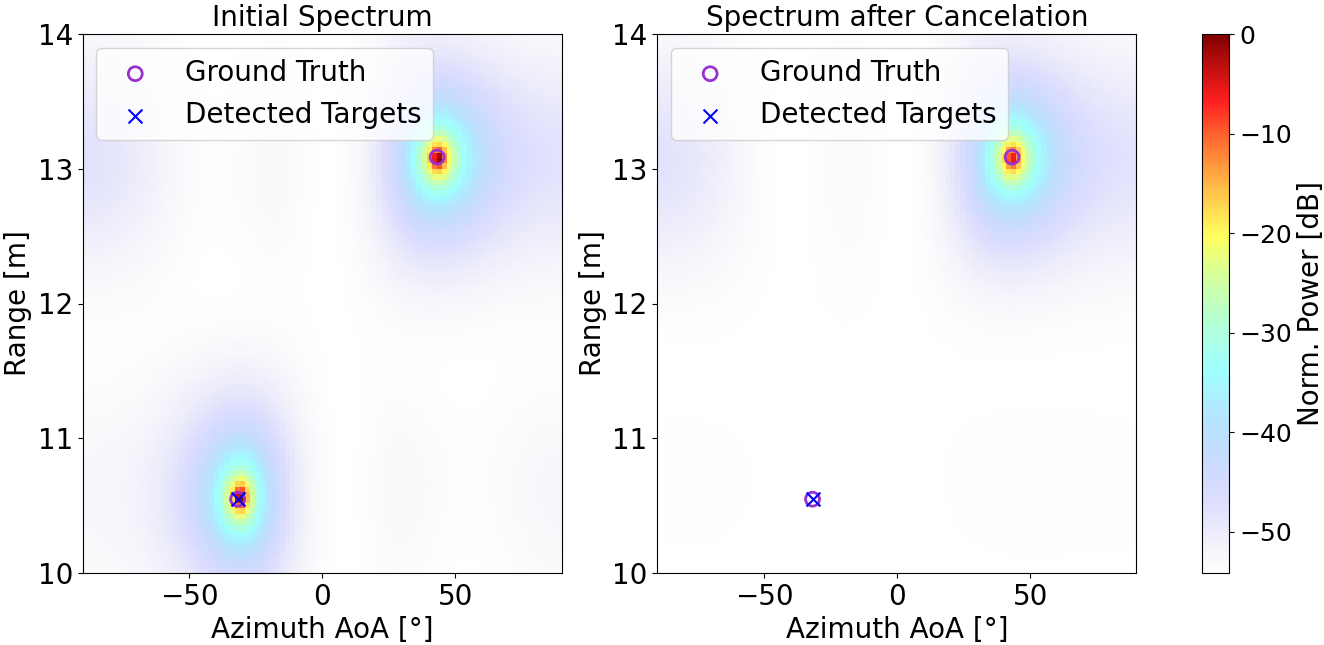}
   \caption{Resolvable targets.}
   \label{fig:spectrum_resolvable}
\end{subfigure}

\vspace{0.2cm}

\begin{subfigure}[b]{0.5\textwidth}
   	\includegraphics[width=\linewidth]{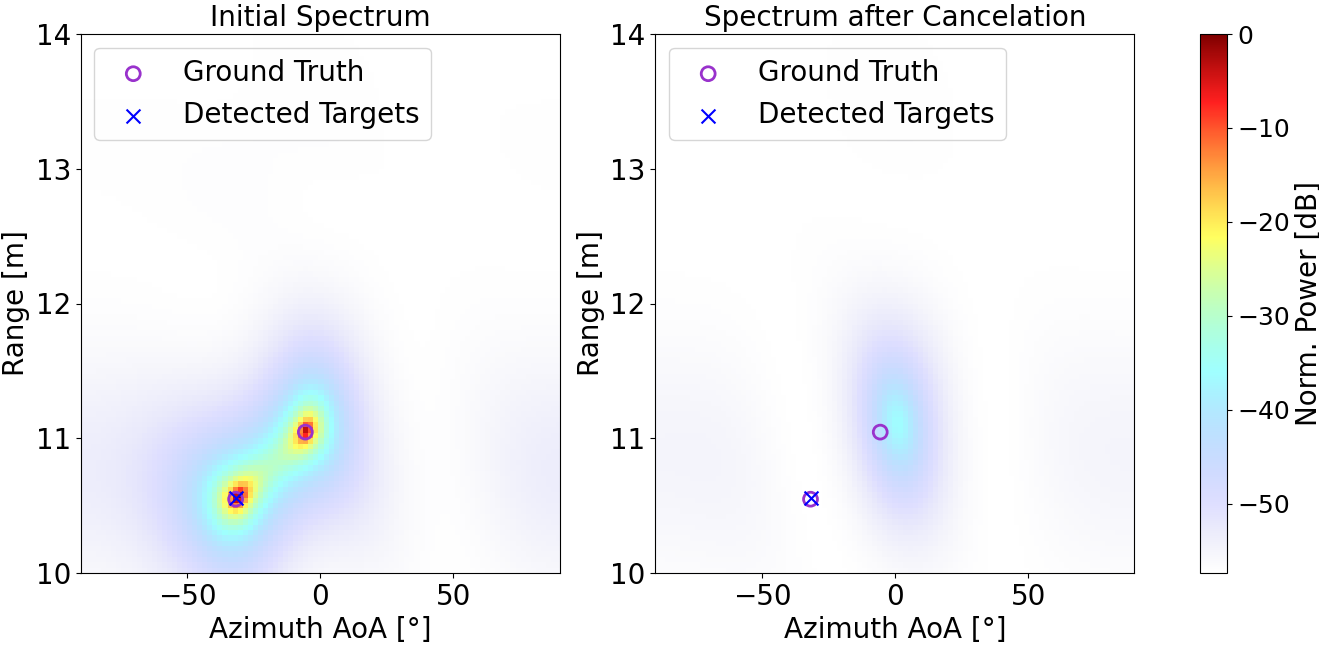}
   \caption{Non-resolvable targets.}
   \label{fig:spectrum_unresolvable}
\end{subfigure}

\caption{Comparison of 2D \gls{music} spectra before (left) and after (right) cancelation of a target. In (a), targets are resolvable, whereas in (b) the residual spectrum is displaced from the true target due to the targets being non-resolvable.}
\label{fig:cancelation}
\end{figure}

To check the \gls{music} spectrum iteratively for remaining targets, the contribution of detected ones should be removed. We construct the channel contribution of the detected target
\begin{align}
\mathbf{c}_{q} &= \tilde{\mathbf{b}}_{\phi_a}(\hat{\theta}_q) \otimes \tilde{\mathbf{a}}_{\phi_f}(\hat{r}_q)
\label{eq:target_cont},	
\end{align} 
where $\hat{\theta}_q$ and $\hat{r}_q$ are the \gls{aoa} and range estimates of $q$. Then, $\mathbf{c}_q$ is orthogonalized \wrt to all eigenvectors spanning $\mathbf{U}_N$
\begin{equation}
\tilde{\mathbf{c}}_q = \mathbf{c}_q - \mathbf{U}_N \mathbf{U}_N^\text{H}\mathbf{c}_q.
\end{equation}
After normalizing the L2-norm $\tilde{\mathbf{c}}_{q, \text{norm}} = \frac{\tilde{\mathbf{c}}_q}{||\tilde{\mathbf{c}}_q||}$, $\tilde{\mathbf{c}}_{q, \text{norm}}$ is added to $\mathbf{U}_N$ to obtain the updated noise subspace $\tilde{\mathbf{U}}_N =\begin{bmatrix}\mathbf{u}_1,\ \mathbf{u}_2,\ \hdots,\ \tilde{\mathbf{c}}_{q, \text{norm}} \end{bmatrix}$ to be used in (\ref{eq:MUSIC}) to re-estimate the spectrum. Note that this subspace tracking approach does not require to re-calculate the \gls{evd} of $\hat{\mathbf{R}}_y$ \cite{subspace_tracking}. Fig. \ref{fig:spectrum_resolvable} shows an example, where the detected target of the initial spectrum on the left (red cross) is canceled out coherently and thus its contribution does not show up in the updated spectrum on the right. However, Fig. \ref{fig:spectrum_unresolvable} shows that if two targets are not resolvable in any dimension, removing the contribution of one target leads to a slight displacement of the remaining target \wrt to its true location. While this peak can still be detected, the displacement degrades the estimation accuracy. Using multiple starting points helps alleviating this issue, as it enables detecting close peaks. Nonetheless, cancelation enhances the detection capabilities, as it facilitates detecting weak peaks after removing the contribution of stronger ones. 

\subsection{Investigated Peak Selection Routines}

Combining the optionalities to either detect one or multiple targets at once and to iteratively remove their contributions, we define three peak selection routines to be investigated.

\fakepar{Single} Only the strongest point ($N_{\text{start}}=1$) of the coarse grid is used as a starting point for the peak search. If this leads to a detection, the contribution of the target is canceled out before re-computing the spectrum for the next iteration.

\fakepar{Multiple} To detect multiple targets at each iteration, the $N_{\text{start}}$ strongest points of the coarse grid are used as starting points. All targets are removed before re-estimating the spectrum.

\fakepar{Off} This mode also considers $N_{\text{start}}>1$, but no coherent target cancelation is performed. Therefore, computation time is saved at the expense of detection capability. 
\\Routines \textit{single} and \textit{multiple} iterate until no targets are left, while \textit{off} only performs a single peak detection iteration.
\section{Results and Discussion}\label{sec:results}

\subsection{1D vs. 2D Estimation}\label{sec:1D_vs_2D}

\begin{figure*}[!htb]
\centering
\begin{subfigure}{0.49\textwidth} 
	\begin{tikzpicture}
		\begin{semilogyaxis}[
			height = 7.2cm,
    		xlabel={Target Range Difference [m]},
    		ylabel={Probability of Missed Detection},
    		ymin=0.001,
		ymax = 1,
		    vasymptote=1.78,
    		legend pos=north west,
		enlargelimits = false,
    		xmajorgrids=true,
    		yminorgrids=true,
    		grid style=dashed,
		mark repeat = {2},
    		legend columns = 2,
    		legend style={at={(0.01,0.14)},anchor=west, font=\scriptsize, inner xsep=0pt, inner ysep=0pt},
		legend cell align={left},
		every axis plot/.append style={thick},
		]

    	\addplot[
   		color=blue,
   		mark=o,
    	mark options={solid}
    	]
    	table {Sim_results/1D_vs_2D/prob_det/2Doff_5};
    	  	
    	\addplot[
   		color=orange,
   		mark=x,
        mark options={solid}
    	]
    	table {Sim_results/1D_vs_2D/prob_det/2Dsingle_5};
    	
    	\addplot[
   		color=green!70!black,
   		mark=triangle,
    	mark options={solid}
    	]
    	table {Sim_results/1D_vs_2D/prob_det/2Dmultiple_5};

	\addplot[
   		color=red,
   		mark=diamond,
    	mark options={solid}
    	]
    	table {Sim_results/1D_vs_2D/prob_det/1Dmultiple_5};

    	\addplot[
   		color=blue,
   		mark=o,
		dotted,
    	mark options={solid}
    	]
    	table {Sim_results/1D_vs_2D/prob_det/2Doff_15};
    	  	
    	\addplot[
   		color=orange,
   		mark=x,
		dotted,
        mark options={solid}
    	]
    	table {Sim_results/1D_vs_2D/prob_det/2Dsingle_15};
    	
    	\addplot[
   		color=green!70!black,
   		mark=triangle,
		dotted,
    	mark options={solid}
    	]
    	table {Sim_results/1D_vs_2D/prob_det/2Dmultiple_15};

	\addplot[
   		color=red,
   		mark=diamond,
   		dotted,
    	mark options={solid}
    	]
    	table {Sim_results/1D_vs_2D/prob_det/1Dmultiple_15};
    	
    	\legend{2D off (5), 2D single (5), 2D mult. (5), 1D mult. (5),
		2D off (15), 2D single (15), 2D mult. (15), 1D mult. (15)}
    
	\end{semilogyaxis}
\node at (6.35,5) {$\Delta r$};
\end{tikzpicture} 
	\caption{Probability of Missed Detection.}
	\label{fig:1D_vs_2D_prob_det}
\end{subfigure}
\begin{subfigure}{0.49\textwidth} 
	\begin{tikzpicture}
		\begin{semilogyaxis}[
			height = 7.2cm,
    		xlabel={Target Range Difference [m]},
    		ylabel={RMSE [m]},
    		ymin=0.009,
		ymax=10,
		yminorticks = true,
    		legend pos=north west,
    	vasymptote=1.78,
		enlargelimits = false,
    		xmajorgrids=true,
    		yminorgrids=true,
    		grid style=dashed,
		mark repeat = {2},
    		legend columns = 1,
    		legend style={at={(0.01,0.27)},anchor=west, font=\scriptsize, inner xsep=0pt, inner ysep=0pt},
		legend cell align={left},
    		every axis plot/.append style={thick}
		]

    	\addplot[
   		color=blue,
   		mark=o,
    	mark options={solid}
    	]
    	table {Sim_results/1D_vs_2D/rmse/2Doff_5};
    	  	
    	\addplot[
   		color=orange,
   		mark=x,
        mark options={solid}
    	]
    	table {Sim_results/1D_vs_2D/rmse/2Dsingle_5};
    	
    	\addplot[
   		color=green!70!black,
   		mark=triangle,
    	mark options={solid}
    	]
    	table {Sim_results/1D_vs_2D/rmse/2Dmultiple_5};

	\addplot[
   		color=red,
   		mark=diamond,
    	mark options={solid}
    	]
    	table {Sim_results/1D_vs_2D/rmse/1Dmultiple_5};

	    	\addplot[
   		color=blue,
   		mark=o,
		dotted,
    	mark options={solid}
    	]
    	table {Sim_results/1D_vs_2D/rmse/2Doff_15};
    	  	
    	\addplot[
   		color=orange,
   		mark=x,
		dotted,
        mark options={solid}
    	]
    	table {Sim_results/1D_vs_2D/rmse/2Dsingle_15};
    	
    	\addplot[
   		color=green!70!black,
   		mark=triangle,
		dotted,
    	mark options={solid}
    	]
    	table {Sim_results/1D_vs_2D/rmse/2Dmultiple_15};

	\addplot[
   		color=red,
   		mark=diamond,
   		dotted,
    	mark options={solid}
    	]
    	table {Sim_results/1D_vs_2D/rmse/1Dmultiple_15};
    	
    	\legend{2D off (5), 2D single (5), 2D mult. (5), 1D mult. (5),
		2D off (15), 2D single (15), 2D mult. (15), 1D mult. (15)}
    
	\end{semilogyaxis}
\node at (6.35,5) {$\Delta r$};
\end{tikzpicture} 
	\caption{Range RMSE.}
	\label{fig:1D_vs_2D_distance_rmse}
\end{subfigure}

\caption{Comparison of missed detection probability and range RMSE performance between 1D and 2D estimation for increasing range differences between two targets at SNRs of 5 dB and 15 dB (in parentheses of legend).} 
\label{fig:1D_vs_2D_plot}
\vspace{-4mm}
\end{figure*}
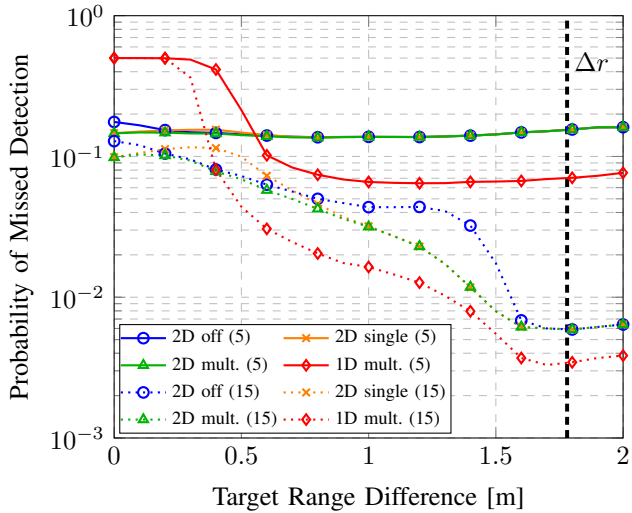
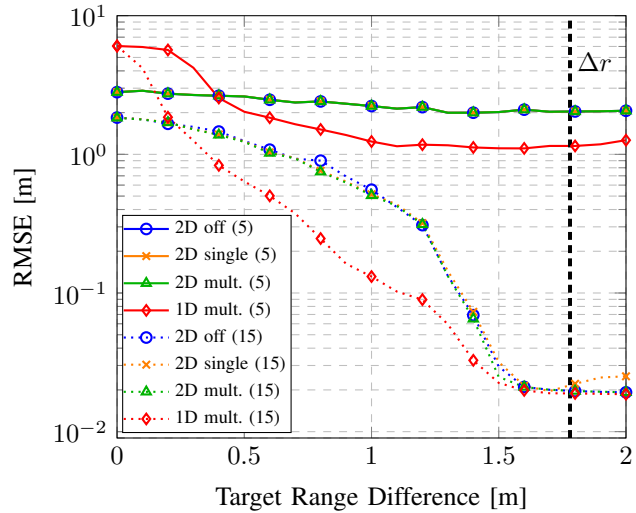

We consider two targets in our simulations as this suffices to show the improved resolution capability using joint estimation. However, the algorithm can be applied to an arbitrary number of targets in practice and future work should include more complex channel models or channels generated by ray tracing tools, \egc \cite{maxray}. The targets are initially placed at $\mathit{J}$ = 10000 random positions resulting in the same range ($< \mathit{r_{\text{max}}}$= 25 m, in line with a typical factory environment) between targets and RX \gls{ula}. From there, one target is incrementally moved away from the RX such that the target range difference is increased by 0.1~m. For each range difference, the same random \glspl{aoa} uniformly drawn between -60° and 60° are kept. The \gls{snr} at RX antenna $k$ is defined as $\text{SNR} = \frac{\mathbb{E}(\mathbf{y}_{k}^2)}{\sigma_{N}^2}$, where $\sigma_{N}^2$ is the noise variance. We consider free-space path loss to model the attenuation of the impulsive scatterers. Note that the fading is modeled through the interference among them. Table~\ref{tab:2D_MUSIC_params} lists the 5G compliant \gls{ofdm} signal specification, the chosen parameters for our proposed algorithm, and the resulting radar characteristics. We consider a 4-antenna \gls{ula} with $\lambda/2$ element spacing and only decimate in the frequency domain (\iec $\mathit{D_a}$ = 1) to avoid spatial aliasing. According to Theorem 1, we choose $A_a = 3$ to allow striding in the spatial domain such that two targets with the same range can still be separated if they are resolvable via their angles.

\begin{table}[h]
	\caption{Simulation Parameters. \label{tab:2D_MUSIC_params}}
	\centering
     \begin{tabular}{|c|c|}
       		\hline
       		Number of subcarriers $\mathit{N}$ & 1500 \\
  			\hline
  			Carrier frequency $\mathit{f_c}$ & 3.5 GHz \\
  			\hline
  			Subcarrier spacing $\Delta f$ & 60 kHz \\
  			\hline
  			Number of RX antennas $\mathit{K}$ & 4 \\
  			\hline
  			Antenna spacing $\mathit{d}$ & $\lambda/2 \approx$ 0.05 m \\
  			\hline
            Frequency aperture $\mathit{A}_{f}$ & 1401 \\
  			\hline
  			Frequency decimation $\mathit{D}_{f}$ & 100 \\
  			\hline 
  			Antenna aperture $\mathit{A}_{a}$ & 3 \\
  			\hline
  			Antenna decimation $\mathit{D}_{a}$ & 1 \\
  			\hline
  			Frequency stride $\mathit{S}_{f}$ & 1 \\
  			\hline
  			Antenna stride $\mathit{S}_{a}$ & 1 \\
  			\hline
  			Starting points for peak search $\mathit{N}_{\text{start}}$ & 10 \\
  			\hline
  			Range resolution $\Delta r$ & 1.78 m \\
  			\hline
  			Unambiguous range $\mathit{r_{\text{max}}}$ & 25 m \\
  			\hline
    \end{tabular}
\end{table}

We define the \gls{rmse} for parameter $\Theta$ (either range $r$ or azimuth \gls{aoa} $\theta$) as
\begin{align}
\text{RMSE} = \sqrt{\frac{1}{2J}\sum_{j=1}^{J} \sum_{q=1}^{2} |\hat{\Theta}_q - \Theta_q|^2},
\label{eq:RMSE}
\end{align}
where $\hat{\Theta}_q$ and ${\Theta}_q$ are estimate and true value of the respective parameter for target $q$. For computing the \gls{rmse}, we put the main focus on the target closer to the \gls{ula} (first target), \iec in case of a single detection we compute the error to the first target. If both targets are detected, the estimate $\mathit{\hat{r}_q}$ that is closer to the \gls{ula} is assigned to the first target and the remaining estimate to the second one. Note that we assign detected targets to true targets only based on the range estimate. To compute errors in case of missed detections, we either use the maximum of the pre-computed coarse grid as the estimate for the first target (in case of zero detections) or remove the contribution of the first detected target and use the maximum of the resulting coarse grid as the estimate for the second one. We discard the 1\% highest and lowest errors for the \gls{rmse} calculation, as outliers can be removed with tracking techniques \cite{5GCAR}.

First, the general advantages of joint estimation are shown by outlining the performance differences of 1D range-only estimation with \gls{music} to our algorithm. For 1D estimation we use the same parameters (Table~\ref{tab:2D_MUSIC_params}), except choosing $A_a = 1$.
\\Fig. \ref{fig:1D_vs_2D_prob_det} shows the probability of missed detection for 1D range-only and 2D estimation at \glspl{snr} of 5~dB and 15~dB. To limit the number of curves we only plot peak selection routine \textit{multiple} for 1D estimation. It can be seen that for both \glspl{snr}, the 2D estimation methods already achieve low probabilities of missed detection for targets with the same range, whereas the 1D range-only estimation can only detect a single target in such cases. Nonetheless, 1D estimation is more robust for higher target range differences, especially at an \gls{snr} of 5~dB. Observing the performance at 15~dB clarifies the benefit of recomputing the spectrum after removing the contributions of detected targets, as \textit{2D off} (no target cancelation) shows the worst performance. Once the targets are fully resolvable in range, the 2D estimation routines attain missed detection probabilities of roughly 0.006, while \textit{1D multiple} performs slightly better and achieves circa 0.004. Overall, \textit{2D multiple} exhibits the best performance of the investigated 2D routines. 

The range \gls{rmse} curves (Fig. \ref{fig:1D_vs_2D_distance_rmse}) show similar progressions. All 2D peak selection routines display a comparable \gls{rmse} performance. At 15~dB, both 1D and 2D techniques converge to error floors of roughly 0.02~m. The problem of target detection is complex to analyze analytically, thus we plot the achievable range resolution $\Delta r$ from Eq. (\ref{eq:range_res}) with our setup as a reference. In general, our algorithms allow to discriminate the targets earlier than the theoretical resolution.
\subsection{Benefits of CSI Decimation}

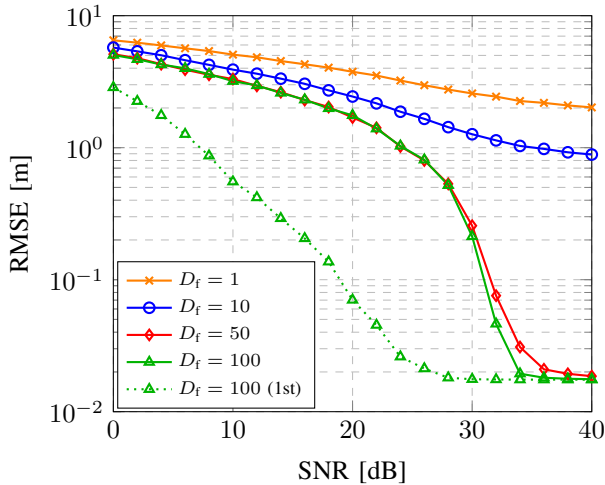
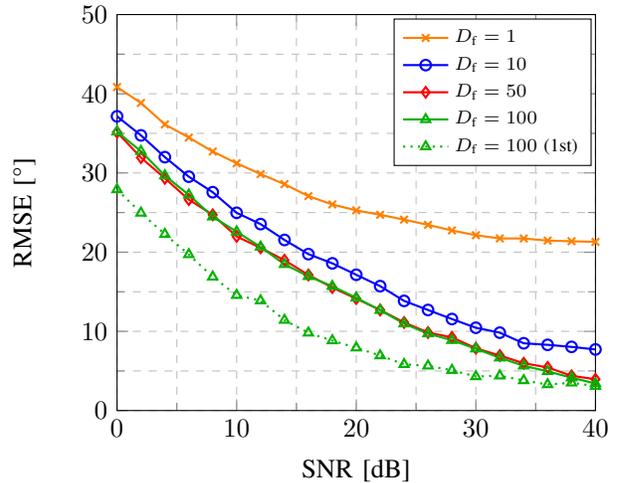
\begin{figure*}[!htb]
\centering
\begin{subfigure}{0.49\textwidth}
      \begin{tikzpicture}
	\begin{semilogyaxis}[
    		height = 6.85cm,
    		xlabel={SNR [dB]},
    		ylabel={RMSE [m]},
    		ymin=0.01, ymax=10,
    		legend pos=north west,
		enlargelimits = false,
    		xmajorgrids=true,
    		yminorgrids=true,
    		grid style=dashed,
    		legend columns = 1,
    		legend style={at={(0.215,0.38)}, anchor=north, font=\scriptsize},
		legend cell align={left},
		every axis plot/.append style={thick}
		]
    	
    	\addplot[
   		color=orange,
   		mark=x,
    	mark options={solid}
    	]
    	table {Sim_results/diff_Df/rmse/distance/1multiple_distance};

	\addplot[
   		color=blue,
   		mark=o,
    	mark options={solid}
    	]
    	table {Sim_results/diff_Df/rmse/distance/10multiple_distance};

	\addplot[
   		color=red,
   		mark=diamond,
    	mark options={solid}
    	]
    	table {Sim_results/diff_Df/rmse/distance/50multiple_distance};
    	
    	\addplot[
   		color=green!70!black,
   		mark=triangle,
    	mark options={solid}
    	]
    	table {Sim_results/diff_Df/rmse/distance/100multiple_distance};
    	
    	\addplot[
    		color=green!70!black,
    		mark = triangle,
		dotted,
    	mark options={solid}
    	]
    	table {Sim_results/diff_Df/rmse/distance/100multiple_distance_target1};
    	
	\legend{$D_\text{f} = 1$, $D_\text{f} = 10$, $D_\text{f} = 50$, $D_\text{f} = 100$, $D_\text{f} = 100$ (1st)}
    	
	\end{semilogyaxis} 
\end{tikzpicture} 
   \caption{Range RMSE.}
   \label{fig:diff_Df_distance_rmse}
\end{subfigure}
\begin{subfigure}{0.49\textwidth}
       \begin{tikzpicture}
	\begin{axis}[
    		height = 6.85cm,
    		xlabel={SNR [dB]},
    		ylabel={RMSE [\textdegree]},
    		ymin=0,
		ymax=50,
		minor tick num=1,
		yminorticks = true,
    		legend pos=north west,
		enlargelimits = false,
		grid = both,
    		grid style=dashed,
    		legend columns = 1,
    		legend style={at={(0.785,0.99)}, anchor=north, font=\scriptsize},
		legend cell align={left},
		every axis plot/.append style={thick}
		]
    	\addplot[
   		color=orange,
   		mark=x,
    	mark options={solid}
    	]
    	table {Sim_results/diff_Df/rmse/azimuth/1multiple_azimuth};

	\addplot[
   		color=blue,
   		mark=o,
    	mark options={solid}
    	]
    	table {Sim_results/diff_Df/rmse/azimuth/10multiple_azimuth};

	\addplot[
   		color=red,
   		mark=diamond,
    	mark options={solid}
    	]
    	table {Sim_results/diff_Df/rmse/azimuth/50multiple_azimuth};
    	
    	\addplot[
   		color=green!70!black,
   		mark=triangle,
    	mark options={solid}
    	]
    	table {Sim_results/diff_Df/rmse/azimuth/100multiple_azimuth};
    	
    	\addplot[
    		color=green!70!black,
    		mark=triangle,
		dotted,
    	mark options={solid}
    	]
    	table {Sim_results/diff_Df/rmse/azimuth/100multiple_azimuth_target1};
    	
	\legend{$D_\text{f} = 1$, $D_\text{f} = 10$, $D_\text{f} = 50$, $D_\text{f} = 100$, $D_\text{f} = 100$ (1st)}
    	
   	\end{axis}    
\end{tikzpicture} 
   \caption{Azimuth RMSE.}
   \label{fig:diff_Df_azimuth_rmse}
\end{subfigure}

\caption{Range and azimuth \gls{rmse} curves for decimations $D_f = 1, D_f = 10, D_f = 50$, and $D_f = 100$ and routine \textit{multiple}. All setups consider sub-arrays with the same number of samples. Hence, higher $D_f$ leads to a higher frequency aperture $A_f$.}
\label{fig:diff_Df_rmse}
\end{figure*}

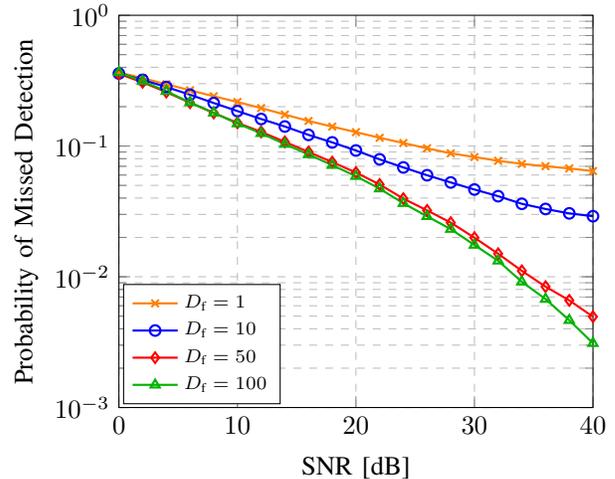
\begin{figure}[!htb]
\centering
\begin{tikzpicture}
	\begin{semilogyaxis}[
			height = 6.8cm,
    		xlabel={SNR [dB]},
    		ylabel={Probability of Missed Detection},
    		ymin=0.001, 
		ymax=1,
    		legend pos=north west,
		enlargelimits = false,
    		xmajorgrids=true,
    		yminorgrids=true,
    		grid style=dashed,
    		legend columns = 1,
    		legend style={at={(0.175,0.315)},anchor=north, font=\scriptsize},
		legend cell align={left},
		every axis plot/.append style={thick}
		]

    	\addplot[
   		color=orange,
   		mark=x,
    	mark options={solid}
    	]
    	table {Sim_results/diff_Df/prob_det/1multiple};

	\addplot[
   		color=blue,
   		mark=o,
    	mark options={solid}
    	]
    	table {Sim_results/diff_Df/prob_det/10multiple};

	\addplot[
   		color=red,
   		mark=diamond,
    	mark options={solid}
    	]
    	table {Sim_results/diff_Df/prob_det/50multiple};
    	
    	\addplot[
   		color=green!70!black,
   		mark=triangle,
    	mark options={solid}
    	]
    	table {Sim_results/diff_Df/prob_det/100multiple};

    	\legend{$D_\text{f} = 1$, $D_\text{f} = 10$, $D_\text{f} = 50$, $D_\text{f} = 100$}
    	\end{semilogyaxis}
\end{tikzpicture} 
\caption{Probability of missed detection for decimations $D_f = 1, D_f = 10, D_f = 50$, and $D_f = 100$ and routine \textit{multiple}. As in Fig. \ref{fig:diff_Df_rmse}, all sub-arrays have the same number of samples.}
\label{fig:diff_Df_prob_det}
\end{figure}

To demonstrate the benefits of decimating in the subcarrier domain, we compare the parametrization in Table \ref{tab:2D_MUSIC_params} against setups with i) $A_f = 15$, $D_f = 1$, ii) $A_f = 141$, $D_f = 10$, and iii) $A_f = 701$, $D_f  = 50$. The number of samples per sub-array is identical ($\mathit{M} = 45$), but the achievable range resolution is reduced by a factor equal to the reduction in $A_f$. To keep the computational effort equal, $L = 200$ sub-arrays were used for all setups. For this experiment we place the two targets at random positions within $r_\text{max}$ without a fixed range difference.
\\Figs. \ref{fig:diff_Df_rmse} and \ref{fig:diff_Df_prob_det} show that a higher range resolution improves both the \gls{rmse} and the missed detection probability performance significantly. Nonetheless, one can observe that a higher \gls{snr} is necessary to achieve similiar missed detection probability and \gls{rmse} capabilities as in Fig. \ref{fig:1D_vs_2D_plot}. This can be explained by trials where the targets are either placed so close that they can not be resolved in either domain, or so far apart that the second target can not be detected due to the path-loss attenuation making them fall below the detection threshold. We therefore as a reference included the curves for the first target only (\textit{1st}), which for $D_f = 100$ (corresponding to \textit{2D multiple} in Fig. \ref{fig:1D_vs_2D_plot}) converge to about 0.01 m and 3°, respectively.

\subsection{Computational Complexity Comparison}

Finally, we demonstrate the savings in computational complexity by comparing the parametrization in Table \ref{tab:2D_MUSIC_params} ($\mathit{D_f} = 100$) with the smoothing technique without decimation ($\mathit{D_f} = 1$) \cite{spotfi}. The setups result in  $M = 45$ and $M = 4203$ elements per sub-array, respectively. The number of \glspl{flop}, \iec complex multiplications and summations, for a single evaluation of (\ref{eq:MUSIC}) is approximated as $2M^2(M-Q)$, thus $O(M^3)$, given $Q$ is small. This leads to $\approx 148$ G\glspl{flop} for $\mathit{D_f} = 1$ and $\approx 174$ k\glspl{flop} for $\mathit{D_f} = 100$. Note that the \gls{evd}'s complexity can also be assumed to be $O(M^3)$, but is only computed once, while (\ref{eq:MUSIC}) is evaluated for all points in the coarse grid, making it the bottleneck. The numbers show that we can reduce the \glspl{flop} by a factor of $\approx 8.5 \cdot 10^{5}$ in a 5G system operating at 3.5 GHz with 90 MHz bandwidth. The gain comes at the price of a reduced unambiguous range, but in typical indoor scenarios the targets' maximum ranges have a ceiling, making this disadvantage tolerable. Moreover, a slight processing gain enhancement due to the higher $M$ can be achieved by not decimating.
\section{Conclusion}\label{sec:conclusion}

We presented an algorithm for joint estimation of range and \gls{aoa} using \gls{music} and \gls{ofdm} signals. \gls{csi} decimation significantly reduces the required number of computations for the considered 5G scenario and makes joint multi-dimensional estimation feasible in practical systems. We demonstrated that our joint approach improves the probability of detection for closely spaced targes compared to 1D range-only estimation, while achieving similar range errors. Moreover, we have proposed efficient single and multi-target detection algorithms.
\section*{Acknowledgments}
The authors gratefully acknowledge helpful discussions with Thorsten Wild, Stephan Saur and Traian Emanuel Abrudan.

\bibliographystyle{IEEEtran}
\bibliography{2D_MUSIC_Main}

\end{document}